\documentclass{ws-fnl}
\usepackage{cite}
\usepackage{amsmath,graphicx}
\usepackage{amsfonts}
\usepackage{amssymb}
\begin{document}

\markboth{Stefano Lepri}{Large Fluctuations in Amplifying Graphs}

\catchline{}{}{}{}{}
%%%%%%%%%%%%%%%%%%%%%%%%%%%%%%%%%%%%%%%%%%%%%%%%%%%%%%%%%%%%%%%%%%%%

\title{Large Fluctuations in Amplifying Graphs}
\author{Stefano Lepri}
\address{Consiglio Nazionale delle Ricerche, Istituto dei Sistemi Complessi, \\
Via Madonna del Piano 10 I-50019 Sesto Fiorentino, Italy\\
stefano.lepri@isc.cnr.it} 
%\address[mysecondaryaddress]{Istituto Nazionale di Fisica Nucleare, Sezione di Firenze, via G. Sansone 1 I-50019, Sesto Fiorentino, Italy}

\maketitle

\begin{history}
\received{(received date)}
\revised{(revised date)}
%\accepted{(Day Month Year)}
%\comby{(xxxxxxxxxx)}
\end{history}

\begin{abstract}
We consider a model for chaotic diffusion with amplification on graphs 
associated with
piecewise-linear maps of the interval [S. Lepri, \textit{Chaos Sol. Fractals}, 
\textbf{139},110003 (2020)]. We determine the conditions for
having fat-tailed invariant measures by considering 
approximate solution of the Perron-Frobenius equation for generic
graphs.
An analogy with the statistical mechanics of a directed
polymer is presented that allows for a physically appealing 
interpretation of the statistical regimes. 
The  connection between non-Gaussian statistics and the generalized 
Lyapunov exponents $L(q)$ is illustrated. Finally, some 
results concerning large graphs are reported.
\vspace{0.2cm}
\end{abstract}

\section{Introduction}

Large fluctuations are one of the distinctive features of complexity, 
being associated to lack of a characteristic scale and to extreme 
events.This work is part of a research program aimed at characterizing 
large fluctuations caused by the joint effect of energy diffusion
and inhomogeneous amplification
or growth.  Diffusion can originate from underlying disorder and 
scattering and/or chaotic motion, while growth stems for external 
energy pumping into the system.  This leads to non-Gaussian fluctuations
of the relevant physical quantities, whose statistical 
distributions can have 
fat-tails,  leading to domination of a 
single event and lack of self-averaging of measurements \cite{bouchaud1990anomalous}.
This is 
well-known for multiplicative stochastic processes 
\cite{sornette1998multiplicative,Garcia-Ojalvo1999}
and chaotic dynamical systems
that display intermittency and multifractality 
\cite{crisanti2012products}. 

A particularly interesting 
form of disorder is the one arising in    
dynamical systems defined on graphs. They
have many fascinating and diverse applications 
to describe complex interacting units with non-uniform connectivity 
\cite{Porter}.  When hetereogeneous reaction is added 
a non trivial interplay between the connectivity  and the local reaction
emerges  \cite{cencetti2018reactive}.

Among the
many possible physical examples, the example we mostly refer to is
the one 
of active, disordered optical media  where light amplification
and 
scattering coexist. \cite{Wiersma2008}. This  
occurs in random lasers where indeed
fat-tailed distributions of emission intensities are
observed experimentally \cite{Lepri2007,Lepri2013,raposo2015analytical, Ignesti2013,Uppu2014,gomes2016observation}.

This work reviews and extends some of the results of 
\cite{lepri2020chaotic} were we introduced 
a simple dynamical system consisting of a map that couples chaotic 
diffusion and energy growth and dissipation. 
Nonlinear maps are time-discrete dynamical models, widely studied to establish 
the emergence of macroscopic behavior from microscopic chaos 
\cite{klages2007microscopic}. 
The model is inspired by experiments on \textit{lasing networks }
\cite{lepri2017complex,giacomelli2019optical}, consisting 
of active (lasing) and passive optical fibers supporting many
optical modes, excited by external pumping.
Optical coupling among the fibers provides a form of topological disorder and 
the system can be considered, loosely speaking, 
as a random laser on a graph.  Another related experimental setup
has been realized with nanophotonic devices by 
coupling a mesh of subwavelength waveguides
\cite{gaio2019nanophotonic}.
Heuristically, one may think of light as a bunch of 
rays undergoing chaotic diffusion and (site-dependent) 
amplification on such graph.
As a matter of fact, the classical dynamics of particles 
on graphs can be described by simple maps. 
Trajectories of a particle on a graph, undergoing scattering at its 
vertices, are in one-to-one correspondence  
with the ones of one-dimensional piecewise chaotic maps \cite{barra2001classical,tanner2000spectral,pakonski2001classical}. 

The plan of the paper is as follows
In Section \ref{sec:model} the recall and extend the map model
introduced \cite{lepri2020chaotic} along with some examples. 
In Section \ref{sec:diffu} we consider an approximate 
equation for the invariant measure and discuss the 
conditions for the appearance of the fat-tailed distributions.
In Section \ref{sec:fat} and examine the 
symmetries of the problem. Such conditions can be recasted 
in terms of a statistical mechanics problem: 
a polymer with a finite number of configurations 
in a random energy landscape, as described in Section
\ref{sec:statmech}. A useful approach is based 
generalized Lyapunov exponents as discussed in Section
\ref{sec:gle}. In Section \ref{sec:2sites} we report 
explicit analytical results for the simplest case of 
a two-sites graph. Finally, we extend the 
analysis to examples of large  
graphs in Section \ref{sec:large}.

\section{Graph with diffusion and amplification}
\label{sec:model}

We consider the following map \cite{lepri2020chaotic}
\begin{equation}
{\begin{cases} 
x_{n+1}=f(x_n) \\ E_{n+1}=g(x_n)E_n
\end{cases}}
\label{model}
\end{equation}
where $g(x)$ is positive
and  $x_n$ belongs to the unit interval.
The function $f$ is piecewise linear and we assume that 
the map is 
chaotic with a Lyapunov exponent $\lambda_1>0$.
The unit interval is partitioned in $N$ 
disjoint intervals $I_j$ of equal lengths $1/N$
and we consider a piece-wise constant gain function
$g$,
$$g(x_n)= g_j \quad \textrm{for} \;x_n\in I_j
$$ 
where the constants $g_j\geq 1$ and $0<g_j< 1$ correspond to 
local amplification or dissipation respectively.  
Thus, the "energy" variable $E_n$ is coupled to $x_n$, 
leading to amplification fluctuations.
Also, the 
sequence of multipliers $g(x_n)$ is in one-to-one
correspondence with the symbolic dynamics of the map
$f$ and has the same time correlation
in time. 
Maps of similar form have been considered 
in the context of on-off intermittency \cite{fujisaka1986intermittency} 
and synchronization transition of two 
piecewise-linear chaotic maps \cite{pikovsky1991symmetry}.

Assuming that the stationary invariant measure $P(x,E)$ of the map 
is uniform in $x$, the Lyapunov exponents $\lambda_{1,2}$ 
are computed straightforwardly
\begin{eqnarray}
&& \lambda_1= \int_0^1 \log |f'(x)| dx,\\
&&\lambda_2=\langle \log(g(x))\rangle=
\frac{1}{N}\sum_j \ln g_j 
\label{lambda12}
\end{eqnarray}
%(Notice that $\lambda_2$ has to be computed for the dynamics \textit{without} the
%reflecting boundaries). 

\begin{figure*}[th] 
\begin{center}
\includegraphics[width=0.25\textwidth,clip]{mapscheme.eps}
\includegraphics[width=0.5\textwidth,clip]{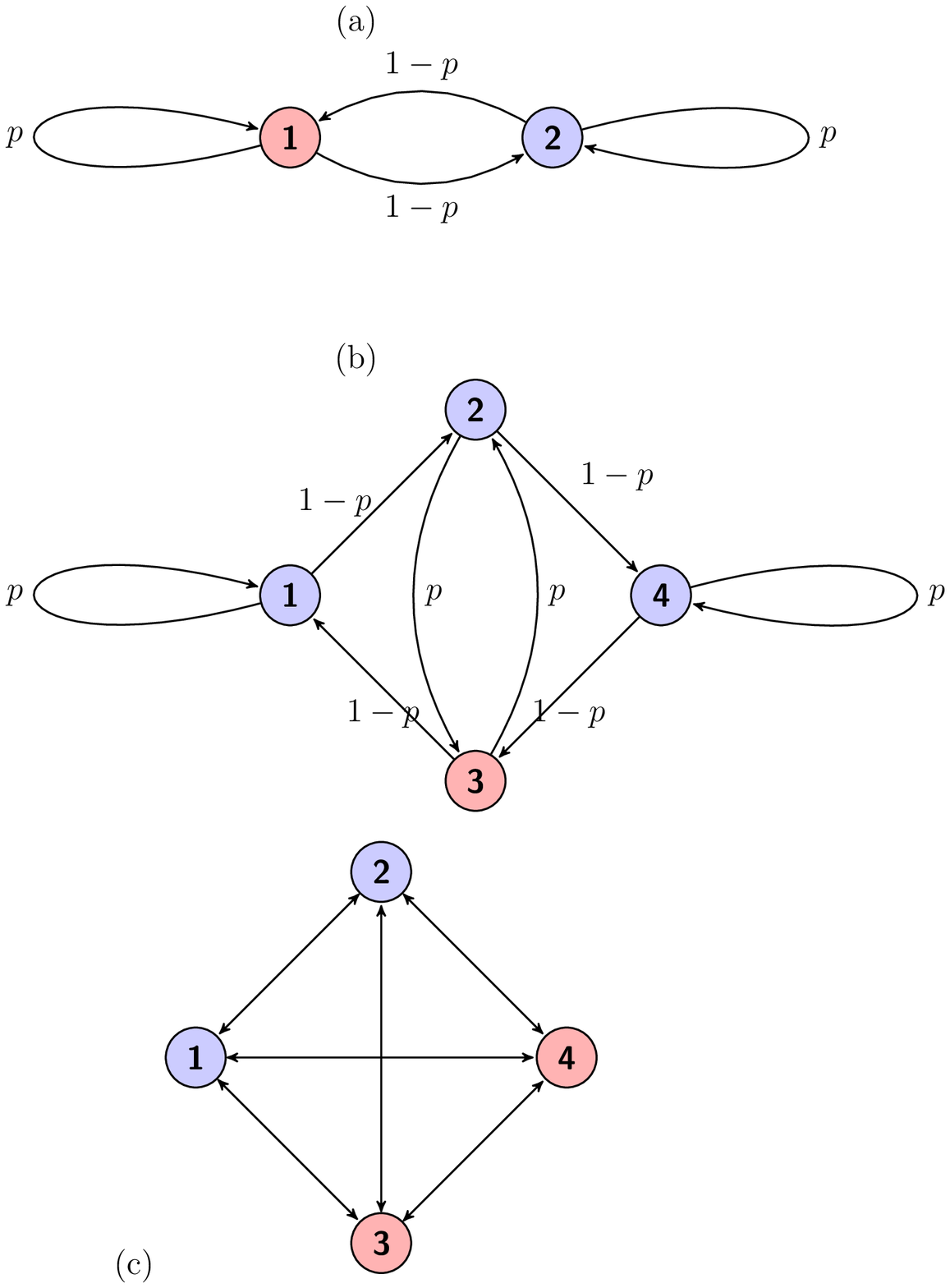}
\end{center}
\caption{Left: three examples of the chaotic map
and (right) their graph representations. 
The analytic expressions are give in the Appendix. 
Red and blue parts represents possible choices 
of amplifying ($g_j>1$) 
and dissipative ($g_j<1$) regions in phase 
space.
}
\label{fig:map}
\end{figure*}

Some specific examples are illustrated in Figure \ref{fig:map} along with
their graph representation, constructed by examining the possible 
transitions in the underlying Markov dynamics.
The first two examples $f_1,f_2$ depend on a parameter 
$p$ (see Appendix) that controls the transition probabilities and 
have the same Lyapunov exponent 
$\lambda_1=-p\log p-(1-p)\log(1-p)$. Note that  $\lambda_1>0$ 
but it is vanishingly small for 
$p$ approaching 0 and 1 where the maps have 
weakly-unstable periodic orbits. 
The third example $f_3$, has $\lambda_1=\log 3$ and corresponds to the 
case of a complete four-sites graph where transition can occur towards any
other site with the same probability. 
This example can be easily extended to arbitrary $N$ (see Section \ref{sec:large}
below). 

In the stable case, $\lambda_2<0$, the orbits tends to be attracted  to the 
origin while $\lambda_2>0$ they  are repelled away and tent to grow 
indefinitely.  
In order to have a bounded invariant measure one needs to
require that the variable $E_n$ neither does drift to infinity
nor is stucked at the origin.  This can be implemented, for 
instance, by assuming that there are
some "barrier" points located at some prescribed values of $E$
(the scale of $E$ is arbitrary).  This can be enforced 
deterministically: for instance for  $\lambda_2<0$ 
setting $E_{n+1}=s$  
when $E_{n}\le 0$,   where 
$s$ is a small positive number. In this way, $E_n$ is stationary 
and ranges in $[0,\infty]$.
The  quantity $s$ is arbitrary, but  
we anticipate that the main results we are interested in 
do not depend on its value.
\footnote{Also setting the variable to a new randomly chosen 
variable $s_n$ will do as well, as long as $s$ is very small as 
it will only affect the shape of the invariant measure close to the 
boundaries \cite{pikovsky1991symmetry}. 
For a discussion of a similar problem for Langevin dynamics see 
Ref. \cite{nakao1998asymptotic} and the bibliography therein.}
In the unstable case, $\lambda_2>0$,  we may impose the 
constraint at, say $E=1$ resetting $E_{n+1}=1 $ whenever  $E_{n}> 1$.
Another possibility would be to use stochastic or determinististic
resetting, or to allow the trajectories to escape,  
see \cite{lepri2020chaotic} for details.

Starting from a "Lagrangian" description in terms of chaotic 
trajectories we can derive the corresponding "Eulerian" 
equations for the probabilities. 
The time-discrete  evolution of the measure $P_n(x,E)$  is 
solution of Perron-Frobenius operator 
\begin{equation}
P_{n+1}(x,E)= 
\sum_j \frac{1}{g_j |f'(y_j)|} P_{n}\left(y_j,\frac{E}{g_j}\right)
\label{eq:fp}
\end{equation}
where $y_j(x)=f^{-1}(x)$ are the $N$ pre-images of $x$. 
Boundary conditions are required  to specify
$P_n(x,E)$
to take properly into account the 
barrier points. 

\begin{figure}[th] 
\begin{center}
\includegraphics[width=0.8\textwidth,clip]{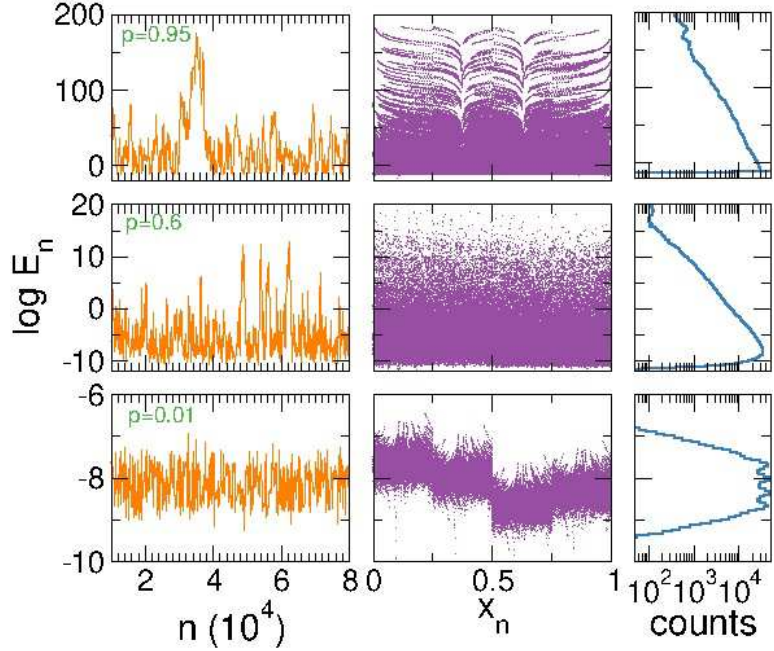} 
\end{center}
\caption{Time evolution and statistics  of iterates of 
the map $f_2$ given in Fig.\ref{fig:map} for  $p=0.95,0.6,0.01$ (bottom to top).
The gain factors are $g_{1,2,4}=0.7$, $g_3=2.7$, corresponding to the Lyapunov
exponent $\lambda_2 \approx -0.0192$. 
Left panels: trajectory snapshots, middle column: distribution of the iterates in phase space
$(x_n,\log E_n)$, histograms of the variable $\log E_n$ in semi-logarithmic scale. }
\label{fig:time}
\end{figure}

To give an idea of the dynamics,  we report in Figure \ref{fig:time}
 some representative trajectories for the map $f_2$ 
 along with the attractors in phase spave and histograms
 of the variable $z=\log E$. Note that an exponential decay 
 at large $z$ is a signature of a power-law tail in the variable
 $E$, that occurs when large fluctuations arise.

\section{Fast chaotic diffusion}
\label{sec:diffu}

Since we are interested in the statistical properties of 
$E$, it is natural to consider, in view of
definition of $g(x)$, the probabilities $P_{j,n}(E)$
to have a particle with energy $E$ in each interval $I_j$.
In general, it is not possible to write a closed equation
for the  $P_{j,n}$ from (\ref{eq:fp}). A simplified case 
is when the Lyapunov exponent $\lambda_1$ is much larger then the
typical rate of change of the energy variable. If so, we may assume that the 
measure becomes rapidly uniform in $x$ within each 
interval $I_j$ .We can thus look for solutions
of (\ref{eq:fp}) independent of $x$,
\begin{equation}
P_{i,n+1}(E)=\sum_j \frac{W_{ij}}{g_j} P_{j,n}\left(\frac{E}{g_j}\right)
\label{pe}
\end{equation}
Then $W$ is the $N\times N$ stochastic matrix for a random walk on 
a $N$-sites, directed graph. In our case 
the transition probabilities are given by the inverses of the map slopes
(see the leftmost part of Figure \ref{fig:map} and the Appendix); $W$ 
is symmetric and doubly stochastic,  $\sum_j W_{j,i}=\sum_j W_{i,j}=1$.
This defines a Markov process in discrete time, as 
better seen by recasting (\ref{pe}) it the 
mathematically equivalent form 
\begin{align*}
&P_{i,n+1}(E)=\sum_j \int  K_{i,j}(E,E')\, P_{j,n}(E') dE'\\
&   K_{i,j}(E,E') = W_{ij} \,  \delta(E-g_j E'),
\end{align*}
that defines the transition rates $K_{ij}(E,E')$ (this last equation
reduces to (\ref{pe}) by integrating the $\delta$s).
We also mention in passing that a Kramers-Moyal expansion, suitable when
all the $g_j$ are close to unity, allows to derive a set of 
coupled Langevin equations for the energies at each graph 
site. The details of the derivation will be reported elsewhere. 
Such equation contain \textit{stochastic advection} terms, 
akin to the one found in the lattice case \cite{lepri2013fluctuations}.

\section{Fat-tailed distributions}
\label{sec:fat}

Let us now examine the possibility of 
having sower-law solutions in the steady state $P_{i,n}(E)=P_i(E)$
of the form $P_i= Q_i E^{-\beta - 1}$;
with $\beta > 0$ for normalizability.
The $Q_i$ are the marginal probabilities to be on site $i$
with whatever energy. 
Substituting this Ansatz in the stationarity condition 
%
%\begin{equation}
% \sum_{j} 
%\left[\frac{W_{ij}}{g_j} P_{j}\left(\frac{E}{g_j}\right)
%-W_{ji}P_{i}(E)\right] =0
%\end{equation}
we obtain a consistency condition for $\beta$
\begin{equation}
Q_i = \sum_j {W_{ij}}  {g_j^\beta} \; Q_j
\label{beta}
\end{equation}
i.e. $Q$  must be an eigenvector of eigenvalue one of the matrix
${W_{ij}}  {g_j^\beta}$, i.e.
\begin{equation}
\det(WG^{\beta}-1)=0,
\label{consist}
\end{equation}
where $G$ is a diagonal matrix having elements $g_j$.
Note that $\beta=0$ is always a solution.
Moreover, the condition is invariant under the transformations
\begin{equation}
g_j \longrightarrow \frac{1}{g_j}, \quad \beta \longrightarrow  -\beta.
\label{symmetry}
\end{equation}
This shoud be interpreted as follows.  If there exist a distribution decaying 
for large $E$  as $E^{-\beta - 1}$ in the stable case  $\lambda_2<0$,  then
the distribution in the unstable case with 
Lyapunov exponent $-\lambda_2$ is $E^{\beta - 1}$ for small $E$
(up to a cutoff set by the upper barrier).

The more interesting regime occurs for $|\beta|<2$ where the measures 
have diverging variance (up to the barrier cutoff). 
Here, we expect a strongly intermittent dynamics,
with dominance of single large fluctuations on the average, as 
in the well-know case of L\'evy-stable distributions
\cite{uchaikin1999}.  
For a fixed $W$, the region in the $N$-dimensional 
parameter space $(g_1,\ldots,g_N)$ where this occurs,
is thus bounded between the hyper-surfaces 
defined by the condition (\ref{consist}) with $\beta=\pm 2$
(see Section \ref{sec:2sites} for an example).

\section{Statistical mechanics analogy} 
\label{sec:statmech}

Let us now show that finding a power-law decay of the 
measure can be interpreted as dual statistical mechanics problem.
First, 
Equation (\ref{consist}) can be rewritten in an equivalent manner by imposing that
the symmetrized matrix $g_i^{\beta/2}{W_{ij}}  g_j^{\beta/2}$ has
an eigenvalue equal to one.
Let us define the quantities $h_j$ and $ E_{ij}$
\[
\ln g_j = \lambda_2 + h_j;\qquad 
%\lambda=\frac{1}{N}\sum_j \ln g_j
E_{ij}(\beta)=-\frac{1}{\beta}\ln W_{ij} -\frac{h_i+h_j}{2}
%\delta_{ij}
\]
with $\sum_j h_j=0$ and $\lambda_2$ is defined by  (\ref{lambda12}).
We can thus rephrase the problem in terms of the statistical mechanics 
of a directed polymer of length $\ell$ whose microscopic configurations are labeled by sequences
of $\sigma_1,\sigma_2 \ldots \sigma_\ell$, with $\sigma_i$ being an integer
assuming values $\sigma_i=1,2\ldots ,N$. The polymer
energy is
\begin{equation}
H=\sum_{i}E_{\sigma_i,\sigma_{i+1}}.
\label{hamil}
\end{equation}
The quantity
$E_{ij}$  thus represents the energy cost between two consecutive
beads of the polymer. It consist of two terms: the one dependent on 
$W$ is a kind of elastic energy, while the $h_j$ represent some 
local energies, akin to the case of the polymer on a 
disordered substrate \cite{krug1997origins}.
Larger positive values of $h_j$ correspond to stronger interaction with the 
substrate itself. 
\footnote{Otherwise it could be represented as a chain of 
$N$-components spins. Here, the spin variables $\sigma_i$ take values in the set $1,2,\ldots N$ on each site
and $h_i$ is a spin-dependent constant magnetic field.
In this interpretation it is reminiscent of the 
Potts model with
nearest-neighbor interactions in 1d. It is different from the simple standard
case where the interaction is of the form $\delta_{\sigma_i,\sigma_{i+1}}$.}
To obtain a thermodynamic state  one has to 
impose some upper and lower bounds to the polymer energy in the same 
way done in the map model.  

The standard approach to compute the partition function 
associated with $H$, is to introduce the 
$N\times N$  transfer matrix
\[
T(\sigma_1,\sigma_2)=\exp[-\beta E_{\sigma_1,\sigma_2}];
\] 
and $\beta$ is interpreted as the inverse temperature .
As it is well known, the partition function of the polymer of length $\ell$
is the trace of $T^\ell$ or,  equivalently,  the sum of the eigenvalues $\tau_i$ of $T$, 
$\sum_i \tau_i^\ell$.
For large $\ell$, we thus have to 
impose that its free energy, namely its the maximal eigenvalue  is 
equal to $\exp(-\beta\lambda_2)$, 
\begin{equation}
\beta \lambda_2 = -\log \tau_1.
\label{maxeig}
\end{equation}
This condition is indeed equivalent to
(\ref{beta}) or  (\ref{consist}).  Notice that, considering
the model parameters as fixed,  this it is a kind of inverse procedure with 
respect to the standard case: one fixes the free energy and wants
to determine the corresponding temperature. 
By virtue of the Perron-Frobenius theorem since the matrix $T$ is strictly positive
then the leading eigenvalue $\tau_1$ is strictly positive and non degenerate.
Also for a finite $N$ it an analytic function of the element so there 
are no phase transitions.

The analogy is also suggestive to understand the difference between
the fat-tailed and Gaussian regimes.  As said,  the first case corresponds 
to $0<\beta<2$.  In the polymer language, this would corrrespond
to the high-temperature regime where the elastic energy terms dominate on 
the pinning terms.  In
other words, the polymer is very stiff and the typical lowest-energy configurations
will be trapped close to the largest $h_j$. These configurations give large 
fluctuations above the average in agreement with the above point of view.  On the contrary, 
for low temperatures the polymer is very loose,  and 
explores the whole configuration space at low cost, making deviations from 
the average behavior very unlikely. In this respect the value 
$\beta=2$ can be consider to define the characteristic temperature where the two
energy terms balance.

Since $\tau_1>0$,  equation (\ref{maxeig}) has no solution for $\lambda_2>0$,
and hence to thermodynamically stable states for the 
polymer problem in the canonical ensemble,  as formulated so far. 
This correspond to the fact that for the dynamical system the 
origin is unstable. To account for this case, one can reason in two
ways.   One is to consider
positive temperature states of a modified Hamiltonian $-H$, 
exploiting the symmetry
(\ref{symmetry}). Alternatively, 
a more suggestive 
thermodynamic interpretation is in terms of \textit{negative absolute 
temperature}.  Let us consider the microcanonical states 
of the polymer with Hamiltonian  $H$ at total energy 
$E$. Than all the microscopic configurations are precisely 
those that reach such such energy. Upon increasing $\lambda_2$ the 
number of such configurations, and thus the 
polymer entropy $S(E)$, should decrease leading to negative temperature 
from the usual relation $\beta=\partial S/\partial E$.
Following the standard reasoning, we can thus regard  
the unstable regime as microcanonical state with negative 
absolute temperature where ensemble equivalence does not hold.
\cite{baldovin2021statistical}.

\section{Generalized Lyapunov exponents}
\label{sec:gle}

A general and elegant approach to look at the problem of fat-tails is
from the point of view of large deviation of Lyapunov 
exponents \cite{pikovsky2016lyapunov}. 
For the dynamical system like (\ref{model}) one can consider 
the generalized Lyapunov exponents $L(q)$, that are 
the growth rates of the $q$th moment of the perturbation
as $\exp(L(q)\tau)$ at large times $\tau$
\cite{crisanti1988generalized,crisanti2012products,pikovsky2016lyapunov}.
The $L(q)$ is the cumulant generating function of the associated 
variable and contains all the information on the fluctuations
beyond the Gaussian regimes  \cite{schomerus2002statistics,zillmer2003multiscaling}. 
The standard Lyapunov exponent is given by $\lambda_2=L'(q=0)$,
and corresponds to the typical average growth of a fluctuation.
Thus,  deviations of $L(q)$ from a linear behavior, $\lambda_2 q$, 
are a signature of  intermittent dynamics \cite{benzi1985characterisation}.
The existence of  power-law stationary tails can be inferred
from inspection of the behavior of the $L(q)$
 \cite{pikovsky1991symmetry,Deutsch1993,Deutsch1994}.
Indeed, if $L(q) > 0$ for large enough $q$ then there is a finite
probability for a small perturbation to grow very large with respect to the average.
More precisely, the condition for power-law distributions with a 
diverging moments for $q>q_*$ is that 
$L(q_*)=0$ \cite{Deutsch1993,pikovsky1991symmetry}.  Such condition must be equivalent to
(\ref{consist}), namely a distribution decaying as $E^{-1-q_*},$ i.e.
$q_*=\beta$.
Also, the  boundaries of the 
region with fat-tailed distribution are defined by  $L(\pm 2)=0$.

For the master equation (\ref{pe}),  the generalized exponents can be computed 
exactly from equations (\ref{moments}).  To this aim, we  consider 
the equation \textit{without barrier boundary conditions} and
consider the moments of $E$ in each portion $I_i$ of the unit 
interval, 
\[
\epsilon^{(q)}_{i,n} \equiv \int E^q P_{i,n}(E)dE .
\]
By multiplying equation (\ref{pe}) by $E^q$ and integrating in $dE$,   
we straightforwardly obtain a set of $N$ difference equations 
\begin{equation}
\epsilon^{(q)}_{i,n+1}=\sum_j W_{ij}  g_j ^q\,\epsilon^{(q)}_{j,n}
\label{moments}
\end{equation}
that are linear and closed at each order (moments of different
order  $q$ are decoupled).
%Considering the average values $q=1$, yields
%the stability condition is that the matrix $WG$ has eigenvalues 
%inside the unit circle.  
Using the same notation as above, 
this amounts simply to compute the largest eigenvalue 
of the matrix $WG^q$ and evaluate $L(q)$ as the logarithm of it, 
a procedure that is basically the same followed to compute the 
cumulant generating function of Markovian dynamics \cite{touchette2009large}.
Note that the matrix can be made symmetric by the same transformation
as given at the beginning of Section \ref{sec:statmech}, so the
eigenvalues are real. Also, comparing with (\ref{symmetry}), we 
see that  the spectrum of generalized exponents is also 
invariant under the transformation $g_j\to 1/g_j$ and $q\to -q$,
which is related to the time-reversal invariance of the trajectories of 
the map.

\begin{figure}[th]
\begin{center}
\includegraphics[width=0.9\textwidth,clip]{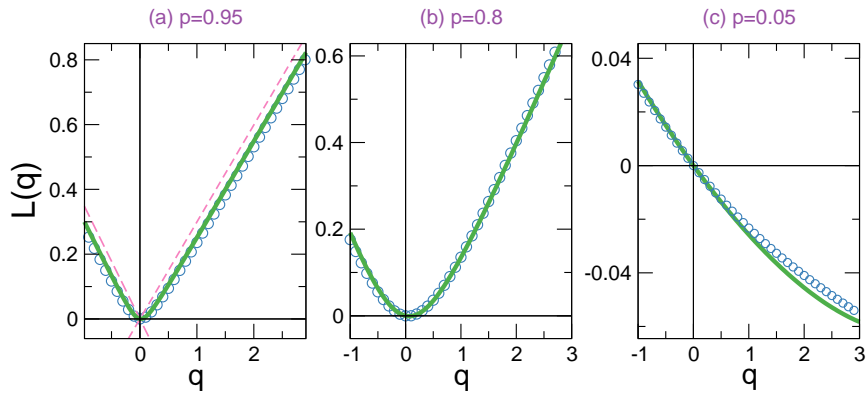} 
\end{center}
\caption{The generalized Lyapunov exponents $L(q)$
for the map $f_2$ and for different values of  parameter $p$ and 
$g=1.2$ $l=0.8$.
Symbols are the numerical values computed 
via the definition, for an ensemble of trajectories 
of finite duration $t=20$ (a,b) and $t=160$ (c); 
solid lines are the $L(q)$ computed as 
the logarithm of the largest
eigenvalue of the matrix $WG^q$ (see text). 
In the case $p=0.95$, we also draw the lines corresponding to the 
maximal and average growth rates,  $q\log l$ and $\lambda_2 q$. 
}
\label{fig:gly}
\end{figure}

For general graphs the eigenvalues can be easily computed numerically.
In Figure \ref{fig:gly} we report the exponents for the case of the 
map $f_2$.  It is known that the exponents are notoriously hard
to compute expecially for large $q$ values that requires sampling
very unlikely trajectories
 \cite{vanneste2010estimating,anteneodo2017importance}.
We thus profit to test the accuracy of the direct method, with respect
to the one based on computation of the eigenvalues .
As seen from the data, for this simple example the direct method
is in reasonable agreement, meaning that sampling accuracy is not
a big issue in those examples.

\begin{figure}[th]
\begin{center}
\includegraphics[width=0.9\textwidth,clip]{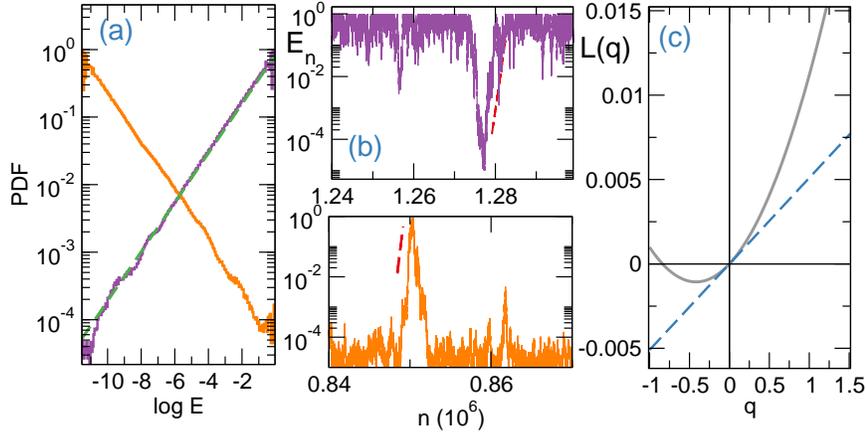} 
\end{center}
\caption{Comparison between the cases with opposite Lyapunov
exponents $\pm \lambda_2$ for the map $f_2$ with
$p=0.5$, $g_1=g_3=g_4=0.9,g_1=1.4$ (orange curves) and 
$g_1=g_3=g_4=1/0.9,g_1=1/1.4$ and (purple).
Panel (a): distributions of $z_n=\log E_n$, the dashed line
is the expected exponential behavior $\exp(-q_*z)$ where 
$q_*=-0.84$ is the given by $L(q_*)=0$. 
Panels (b): time series of $E_n$ showing the build and 
decay of a large fluctuation: the dashed lines correspond 
to exponential growth/decay according to $\exp(\lambda_2 t)$
with $\lambda_2=5.09 \, 10^{-3}$.
Panel (c): the generalized Lyapunov exponents $L(q)$
(solid line) for the case $\lambda_2>0$, the dashed
line is $\lambda_2 q$. 
}
\label{fig:posneg}
\end{figure}

This approach is  
also accurate in reproducing the measured exponents $\pm \lambda_2$.
For instance, as a numerical test for the case $p=0.6$ of 
Figure \ref{fig:posneg}, the condition $L(q_*)=0$ yields $q_*\approx 0.225..$ 
to be compared from the 
fit of the distribution of $z=\log E$ yielding $\exp(-0.223 z)$
for large $z$ (see \cite{lepri2020chaotic} for further numerical
checks).

%It can be checked from equation (\ref{gle}).
%\begin{align}
%&P_{1}(E)=
%\frac{1}{l}{P_{2}\left(\frac{E}{l}\right) } \nonumber\\
%&P_{2}(E)=
% {1\over g  }P_{1}\left({E\over g }\right)
%  \label{boh}
%\end{align}
%Which would suggest the condition $gl=1$

In Figure \ref{fig:posneg}a we compare two cases having parameters
$g_j$ and $1/g_j$ and thus opposite Lyapunov exponents. 
According to (\ref{symmetry}) the statistics of should have 
opposite rates $\exp(\pm q_* z)$, as well verified by the data.  It is also seen that 
the  large fluctuation has a form of statistical symmetry, in the sense
that their shape $E_n$ for $\lambda_2 <0$ would be similar
to $1-E_n$  for $\lambda_2 <0$. (see Figure \ref{fig:posneg}b)
Moreover, rise and fall rates are accurately 
predicted by $L'(q_*)$ and $\lambda_2$, respectively,
as read from Figure \ref{fig:posneg}c.

\section{Two-sites graph: analytical solutions}
\label{sec:2sites}

For the simplest case of the two-sites
graph, corresponding to the map $f_1$ in Figure \ref{fig:map} some 
explicit analytical results can be worked out.
The stationary measure is solution of (letting $g_1=g$, $g_2=l$).
\begin{align}
&P_{1}(E)=
\frac{p}{g} P_{1}\left(\frac{E}{g}\right)+
\frac{1-p}{l}{P_{2}\left(\frac{E}{l}\right) } \nonumber\\
&P_{2}(E)=
 {(1-p) \over g  }P_{1}\left({E\over g  }\right)
  +{p \over l}P_{2}\left({E\over l}\right).
  \label{master}
\end{align}
%\begin{equation*}
%W\equiv
%\begin{pmatrix}
% p & 1-p \\  1-p & p
%\end{pmatrix},\;
%G^{\beta}\equiv 
%\begin{pmatrix}
% g^{\beta} & 0 \\  0 & l^{\beta} 
%\end{pmatrix} 
%\end{equation*}

\begin{itemize}
\item The consistency conditions, 
Eq. (\ref{consist}) yields
\begin{equation}
\det(WG^{\beta}-1)=-p(g^\beta + l^\beta)+(2p-1)g^\beta l^\beta +1=0
\label{consist2sites}
\end{equation}
where the transition matrix is given by $W_1$ in (\ref{matrixW})
and $G^{\beta}\equiv 
\begin{pmatrix}
g^{\beta} & 0 \\  0 & l^{\beta} 
\end{pmatrix}.$
The region in the $(g,l)$ with large fluctuations $|\beta|<2$
is bounded between the curves defined by (\ref{consist2sites})
with $\beta=\pm 2$.

\item 
The statistical mechanics analogy can be worked out explicitely in the 
"spin" interpretation as an Ising chain.
Let $g=\exp(\lambda_2+h)$, $l=\exp(\lambda_2-h)$ in 
(\ref{consist2sites}),  the transfer matrix concides with the 
well-known textbook expression of the 
one-dimensional Ising model with $\beta$ dependent 
parameters
\[
H= \sum_i[-J\sigma_i\sigma_{i+1}+h\sigma_i]
\]  
%are 
%easily computed, 
%
%yielding
($\lambda_2\equiv \lambda$).
%\[
%\exp(\beta\lambda)=
%\frac{p\cosh \beta h \pm \sqrt{p^2 \sinh^2\beta h +(p-1)^2}}{2p-1}
%\]
%\[
%\exp(-\beta\lambda)=\tau_1=
% p\cosh \beta h + \sqrt{p^2 \sinh^2\beta h +(p-1)^2}
%\]
Upon letting 
\[
p=\frac{e^{\beta J}}{e^{\beta J} + e^{-\beta J}};\quad
1-p=\frac{e^{-\beta J}}{e^{\beta J} + e^{-\beta J}};\quad
J(\beta) = \frac{1}{2\beta}\ln (\frac{p}{1-p})
\]
condition (\ref{maxeig}) is rewritten in the familiar form
\[
\exp(-\beta\lambda)=
 e^{\beta J}\cosh \beta h + \sqrt{e^{2\beta J} \sinh^2\beta h +e^{-2\beta J}}
\]
%with $e_0(\beta)=\ln(2\cosh \beta J)/\beta$. 
%Thus $\lambda$ is the 
%free energy of a 
%the condition amount to fix the temperature of the Ising such as the free
%energy is $\lambda$ (remember $\lambda<0$). 
Also, it gives a nice 
interpretation of the parameter $p$. The very definition of $J$  makes 
transparent that $p$ is interpreted as a  probability of a spin-flip 
and controls the type of interaction. In particular:
\begin{itemize}
\item for $p=1/2$: $J=0$ and
\[
\beta\lambda= -\ln \cosh \beta h 
\]
that correspond to 1d Ising paramagnet in external field $h$
\item For $1/2<p<1$: $J>0$ ferromagnetic interaction;

\item For $0<p<1/2$: $J<0$ antiferromagnetic interaction.
\end{itemize}
In this language, the variable of interest is the 
magnetic energy of the spin chain.   
\item Generalized Lyapunov exponents  can be computed 
analytically as described above yielding \cite{lepri2020chaotic} 
\begin{equation}
L(q) = \log 
\left|\frac{p(g^q + l^q)+
\sqrt{p^2(g^q + l^q)^2-4(2p-1)g^q l^q} }
{2}\right|.
\label{gle}
\end{equation}
and it can be checked that the condition $L(q_*)=0$ yields
the same as (\ref{consist2sites}).
\end{itemize}

\begin{figure}[th] 
\begin{center}
\includegraphics[width=0.5\textwidth,clip]{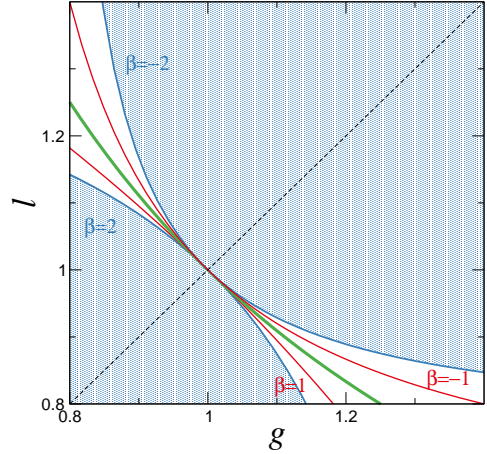} 
\end{center}
\caption{Phase diagram in the parameter plane $(g,l)$
for two-sites graph, corresponding to the map $f_1$ in 
Figure \ref{fig:map} with $p=0.6$. The green line is the bifurcation
line where $\lambda_2=0$. There is an obvious 
reflection symmetry 
around the line $g=l$ (black dashed). 
Shaded blue regions correspond to
finite variance of the variable $E_n$ yielding 
Gaussian fluctuations. The region bounded between the curves 
with $\beta=\pm 2$ is where L\'evy-like fluctuations 
are expected.
 }
 \label{fig:diag}
\end{figure}

In Fig. \ref{fig:diag} we summarize the various statistical 
regimes of the model, distinguishing the parameters values 
where fluctuations have diverging variance.

\section{Large graphs}
\label{sec:large}

So far we have considered graphs with a small number of sites.
A natural question would be how the result change upon 
increasing $N$, in particular whether the fat-tailed regimes
persist. This is not an obvious question. For instance 
in large chaotic systems, the generalized Lyapunov exponents
may become proportional to $q$. The heuristic explanation
is that, due to fast correlation decay in spatio-temporal
chaos, the norm vector is the sum of entries that grow 
almost independently \cite{crisanti1988generalized}.
In the present example however the multipliers $g_j$ are 
quenched and the situation may be different.

\begin{figure}[th] 
\begin{center}
\includegraphics[width=0.7\textwidth,clip]{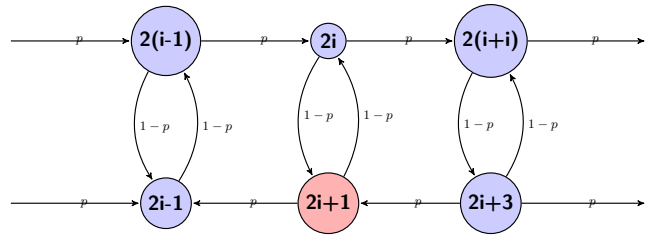} 
\end{center}
\caption{The ladder graph described by the transition matrix (\ref{ladder}).
Configuration with only one active site, labeled by red color.}
 \label{fig:chain}
\end{figure}

For simplicity, let us discuss directly the Markovian dynamics,  Eq. (\ref{pe}).
As a first instance, let us consider the ladder graph composed of $N=2M$ sites 
depicted in Fig. \ref{fig:chain}. For convention, we label the upper
sites with even integers and the lower by odd ones.
The transition matrix has non-zero 
elements given by
\begin{equation}
W_{2i,2(i+1)} = W_{2i+1,2i-1} =p; \qquad W_{2i,2i+1}=W_{2i+1,2i}=1-p
\label{ladder}
\end{equation}
for $i=1\ldots M$, and $W_{ij}=0$ otherwise. Periodic boundaries have 
been assumed.
This is a geometry which can be seen as 
an extension of the map $f_2$ discussed above. 

Another example is the complete graph
\begin{equation}
W_{i,i} = 0; \qquad W_{i,j}=\frac{1}{N-1}\qquad i\neq j
\label{complete}
\end{equation}
which is depicted in the lowest panel of Fig.  \ref{fig:map}
for $N=4$.
For simplicity, let us consider also the case of a single active site with 
all the other having 
equal dissipation, namely $g_j=g>1$ for a certain $j=j_0$ 
and $g_j=l<1$ otherwise. 
With such choice, the Lyapunov exponent is $\lambda_2=(1-1/N)\log l+(\log g)/N $
in both examples, and  approaches the constant value  $\lambda_2\approx \log l<0$ for 
$N$ large.

\begin{figure}[th] 
\begin{center}
\includegraphics[width=1.\textwidth,clip]{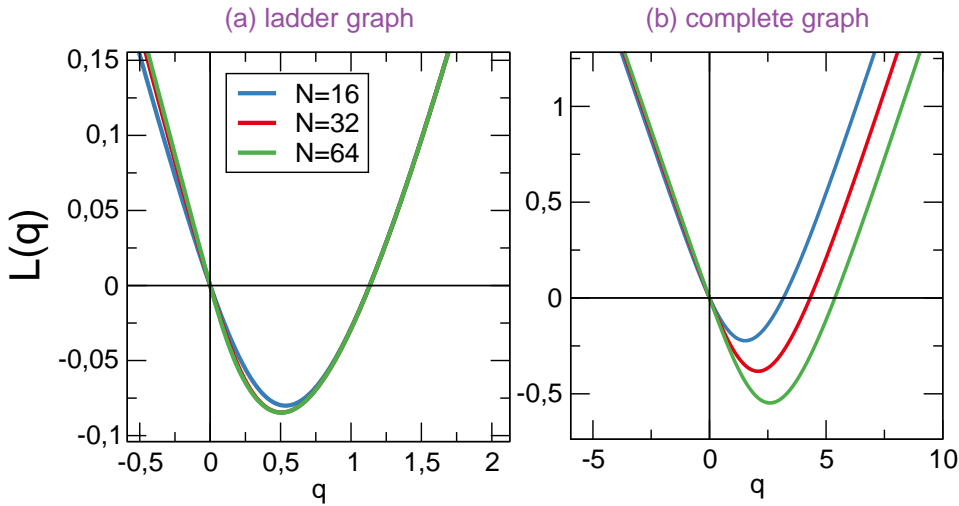} 
\end{center}
\caption{The generalized Lyapunov exponents $L(q)$
for (a) the ladder graphs and (b) the complete graph
and different
number of sites $N$; $p=0.4$ $g_3=3.0$ and $g_j=0.7$ for $j\neq 3$.
}
 \label{fig:glyN}
\end{figure}

In Fig. \ref{fig:glyN} we compare the generalized Lyapunov exponents 
(computed as above) for the two graphs for increasing size $N$.
In the case of the ladder,  the  $L(q)$ are $N$-independent
and the example shown predicts that a fat-tail  with $q_*\approx 1$ should persist
upon increasing the size.
On the contrary for the complete graph $q_*$ grows with $N$ suggesting that
the statistics should turn Gaussian for large enough $N$.

This is qualitatively consistent with the polymer interpretation, 
Equation (\ref{hamil}). In the 
ladder case the elastic energy of the polymer is independent of the 
size. On the contrary for the complete graph the elastic energy decreases
with $N$ making the polymer more and more loose and 
hindering the observation of configurations associated
with  large energy fluctuations.

\section{Conclusion}

Motivated by experiments on active, disordered optical systems , 
we have studied a map model that combines chaotic diffusion and amplification
on a graph \cite{lepri2020chaotic}. 
Within a stochastic approximation  of the 
dynamics, 
given by the Markov process described by (\ref{pe}), 
 we established the conditions for its invariant measure to 
 display fat-tailed distributions in some regime of 
parameters.  
We mostly discussed some specific 
small graphs ($N=2,4$) examples  and extended the results to 
large graphs (ladder and complete).
The model has symmetries that allows to 
consider the stable and unstable cases in a simple way. 
Also, the problem can be interpreted in statistical mechanics language,
an analogy that can be fruitful to interpret the dynamical regimes.
We have confirmed that the Generalized Lyapunov exponents 
provide a useful and simple tool to predict 
the fluctuation statistics and the anatomy of a large 
fluctuation, both in the 
stable and unstable cases. 
The model has its own interest but is also a guidance for interpretration
of experiments in the optical disordered media 
as for instance the lasing networks \cite{giacomelli2019optical}.

%\section*{Declaration of competing interest}
%The author declares no known competing 
%financial interests or personal relationships that could have appeared to
%influence the work reported in this paper.
%
%\section*{Data availability}
%
%Data will be made available on request

\section*{Acknowledgements}

The Author acknowledges S. Iubini, A. Politi and P. Politi for useful discussions and 
A. Di Garbo and R. Mannella for the invitation to the workshop DCP22 
in Pisa.

\section*{Appendix}
For reference we give here the functional forms of the examples
considered in Fig. \ref{fig:map}:
\begin{equation}
f_1(x)=
\begin{cases}
{1 \over p }x & 0\leq x\leq p/2\\
{1 \over 1-p}x+{1-2p \over 2(1-p)} & p/2<x\leq 1/2 \\ 
{1 \over 1-p}x-{1 \over 2(1-p)} & 1/2<x\leq 1-p/2 \\
{1 \over p}x +1 -\frac{1}{p}& 1-p/2<x\leq 1 
\end{cases}
\label{f1}
\end{equation}

\begin{equation}
f_2(x)=
\begin{cases}
x/p          &x<p/4\\
(x-p/4)/(1-p)+1/4  & p/4 \le x < 1/4\\
(x - 1/4)/p+1/2          & 1/4 <x < 1/4+p/4\\
(x - 1/4-p/4)/(1-p)+3/4 & 1/4+p/4 <x < 1/2\\
(x-1/2)/(1-p)     &    1/2<x < 3/4-p/4\\
(x-3/4+p/4)/p + 1/4    &  3/4-p/4< x < 3/4\\
(x - 3/4)/(1-p)+1/2     & 3/4 <x < 1-p/4\\
(x - 1+p/4)/p+3/4   &     1>x > 1-p/4
\end{cases}
\label{f2}
\end{equation}
where $0\le p\le 1$.
The stochastic matrices used in the text are
\begin{equation}
W_1\equiv
\begin{pmatrix}
 p & 1-p \\  1-p & p
\end{pmatrix}
, \quad W_2\equiv
\begin{pmatrix}
p &  1-p&  0 & 0 \\
0 &  0 &  p& 1-p \\
1-p &  p &  0 & 0 \\
0 &  0&  1-p& p \\
\end{pmatrix}
\label{matrixW}
\end{equation}
while for $f_3$ is given by (\ref{complete}) for $N=4$.

\bibliography{mbiblio}

\begin{thebibliography}{10}
\newcommand{\enquote}[1]{``#1''}

\bibitem{bouchaud1990anomalous}
J.-P. Bouchaud and A.~Georges, \enquote{Anomalous diffusion in disordered
  media: statistical mechanisms, models and physical applications},
  \emph{Physics reports} \textbf{195} (1990) 127--293.

\bibitem{sornette1998multiplicative}
D.~Sornette, \enquote{Multiplicative processes and power laws}, \emph{Physical
  Review E} \textbf{57} (1998) 4811.

\bibitem{Garcia-Ojalvo1999}
J.~Garc{\'\i}a-Ojalvo and J.~Sancho, \emph{Noise in spatially extended systems}
  (Springer Verlag, 1999).

\bibitem{crisanti2012products}
A.~Crisanti, G.~Paladin and A.~Vulpiani, \emph{Products of random matrices in
  statistical physics}, volume 104 (Springer Science \& Business Media, 2012).

\bibitem{Porter}
M.~A. Porter and J.~P. Gleeson, \emph{Dynamical Systems on Networks: A
  Tutorial} (Springer series Frontiers in Applied Dynamical Systems: Reviews
  and Tutorials, Switzerland, 2016).

\bibitem{cencetti2018reactive}
G.~Cencetti, F.~Battiston, D.~Fanelli and V.~Latora, \enquote{Reactive random
  walkers on complex networks}, \emph{Physical Review E} \textbf{98} (2018)
  052302.

\bibitem{Wiersma2008}
D.~S. Wiersma, \enquote{The physics and applications of random lasers},
  \emph{Nat. Phys.} \textbf{4} (2008) 359--367.

\bibitem{Lepri2007}
S.~Lepri, S.~Cavalieri, G.~Oppo and D.~S. Wiersma, \enquote{Statistical regimes
  of random laser fluctuations}, \emph{Phys. Rev. A} \textbf{75} (2007) 063820.

\bibitem{Lepri2013}
S.~Lepri, \enquote{Fluctuations in a diffusive medium with gain}, \emph{Phys.
  Rev. Lett.} \textbf{110} (2013) 230603.

\bibitem{raposo2015analytical}
E.~Raposo and A.~Gomes, \enquote{Analytical solution for the {L{\'e}vy}-like
  steady-state distribution of intensities in random lasers}, \emph{Phys. Rev.
  A} \textbf{91} (2015) 043827.

\bibitem{Ignesti2013}
E.~Ignesti, F.~Tommasi, L.~Fini, S.~Lepri, V.~Radhalakshmi, D.~Wiersma and
  S.~Cavalieri, \enquote{Experimental and theoretical investigation of
  statistical regimes in random laser emission}, \emph{Phys. Rev. A}
  \textbf{88} (2013) 033820.

\bibitem{Uppu2014}
R.~Uppu and S.~Mujumdar, \enquote{L\'evy exponents as universal identifiers of
  threshold and criticality in random lasers}, \emph{Phys. Rev. A} \textbf{90}
  (2014) 025801.

\bibitem{gomes2016observation}
A.~S. Gomes, E.~P. Raposo, A.~L. Moura, S.~I. Fewo, P.~I. Pincheira, V.~Jerez,
  L.~J. Maia and C.~B. De~Ara{\'u}jo, \enquote{Observation of {L{\'e}vy}
  distribution and replica symmetry breaking in random lasers from a single set
  of measurements}, \emph{Scientific reports} \textbf{6} (2016) 27987.

\bibitem{lepri2020chaotic}
S.~Lepri, \enquote{Chaotic fluctuations in graphs with amplification},
  \emph{Chaos, Solitons \& Fractals} \textbf{139} (2020) 110003.

\bibitem{klages2007microscopic}
R.~Klages, \emph{Microscopic chaos, fractals and transport in nonequilibrium
  statistical mechanics}, volume~24 (World Scientific, 2007).

\bibitem{lepri2017complex}
S.~Lepri, C.~Trono and G.~Giacomelli, \enquote{Complex active optical networks
  as a new laser concept}, \emph{Physical review letters} \textbf{118} (2017)
  123901.

\bibitem{giacomelli2019optical}
G.~Giacomelli, S.~Lepri and C.~Trono, \enquote{Optical networks as complex
  lasers}, \emph{Physical Review A} \textbf{99} (2019) 023841.

\bibitem{gaio2019nanophotonic}
M.~Gaio, D.~Saxena, J.~Bertolotti, D.~Pisignano, A.~Camposeo and R.~Sapienza,
  \enquote{A nanophotonic laser on a graph}, \emph{Nature communications}
  \textbf{10} (2019) 1--7.

\bibitem{barra2001classical}
F.~Barra and P.~Gaspard, \enquote{Classical dynamics on graphs}, \emph{Phys.
  Rev. E} \textbf{63} (2001) 066215.

\bibitem{tanner2000spectral}
G.~Tanner, \enquote{Spectral statistics for unitary transfer matrices of binary
  graphs}, \emph{Journal of Physics A: Mathematical and General} \textbf{33}
  (2000) 3567.

\bibitem{pakonski2001classical}
P.~Pakonski, K.~Zyczkowski and M.~Kus, \enquote{Classical 1d maps, quantum
  graphs and ensembles of unitary matrices}, \emph{J. Phys. A} \textbf{34}
  (2001) 9303.

\bibitem{fujisaka1986intermittency}
H.~Fujisaka, H.~Ishii, M.~Inoue and T.~Yamada, \enquote{Intermittency caused by
  chaotic modulation. ii: Lyapunov exponent, fractal structure and power
  spectrum}, \emph{Progress of theoretical physics} \textbf{76} (1986)
  1198--1209.

\bibitem{pikovsky1991symmetry}
A.~S. Pikovsky and P.~Grassberger, \enquote{Symmetry breaking bifurcation for
  coupled chaotic attractors}, \emph{Journal of Physics A: Mathematical and
  General} \textbf{24} (1991) 4587.

\bibitem{nakao1998asymptotic}
H.~Nakao, \enquote{Asymptotic power law of moments in a random multiplicative
  process with weak additive noise}, \emph{Physical Review E} \textbf{58}
  (1998) 1591.

\bibitem{lepri2013fluctuations}
S.~Lepri, \enquote{Fluctuations in a diffusive medium with gain},
  \emph{Physical review letters} \textbf{110} (2013) 230603.

\bibitem{uchaikin1999}
V.~V. Uchaikin and V.~M. Zolotarev, \emph{Chance and stability: stable
  distributions and their applications} (Walter de Gruyter, 1999).

\bibitem{krug1997origins}
J.~Krug, \enquote{Origins of scale invariance in growth processes},
  \emph{Advances in Physics} \textbf{46} (1997) 139--282.

\bibitem{baldovin2021statistical}
M.~Baldovin, S.~Iubini, R.~Livi and A.~Vulpiani, \enquote{Statistical mechanics
  of systems with negative temperature}, \emph{Physics Reports} \textbf{923}
  (2021) 1--50.

\bibitem{touchette2018introduction}
H.~Touchette, \enquote{Introduction to dynamical large deviations of {Markov}
  processes}, \emph{Physica A: Statistical Mechanics and its Applications}
  \textbf{504} (2018) 5--19.

\bibitem{crisanti1988generalized}
A.~Crisanti, G.~Paladin and A.~Vulpiani, \enquote{Generalized {Lyapunov
  exponents} in high-dimensional chaotic dynamics and products of large random
  matrices}, \emph{Journal of statistical physics} \textbf{53} (1988) 583--601.

\bibitem{pikovsky2016lyapunov}
A.~Pikovsky and A.~Politi, \emph{Lyapunov exponents: a tool to explore complex
  dynamics} (Cambridge University Press, 2016).

\bibitem{schomerus2002statistics}
H.~Schomerus and M.~Titov, \enquote{Statistics of finite-time {Lyapunov}
  exponents in a random time-dependent potential}, \emph{Physical Review E}
  \textbf{66} (2002) 066207.

\bibitem{zillmer2003multiscaling}
R.~Zillmer and A.~Pikovsky, \enquote{Multiscaling of noise-induced parametric
  instability}, \emph{Physical Review E} \textbf{67} (2003) 061117.

\bibitem{benzi1985characterisation}
R.~Benzi, G.~Paladin, G.~Parisi and A.~Vulpiani, \enquote{Characterisation of
  intermittency in chaotic systems}, \emph{Journal of Physics A: Mathematical
  and General} \textbf{18} (1985) 2157.

\bibitem{Deutsch1993}
J.~M. Deutsch, \enquote{{Generic behavior in linear systems with multiplicative
  noise}}, \emph{Phys. Rev. E} \textbf{48} (1993) 4179--4182.

\bibitem{Deutsch1994}
J.~M. Deutsch, \enquote{{Probability distributions for multicomponent systems
  with multiplicative noise}}, \emph{Physica A} \textbf{208} (1994) 445--461.

\bibitem{touchette2009large}
H.~Touchette, \enquote{The large deviation approach to statistical mechanics},
  \emph{Physics Reports} \textbf{478} (2009) 1--69.

\bibitem{vanneste2010estimating}
J.~Vanneste, \enquote{Estimating generalized {Lyapunov exponents} for products
  of random matrices}, \emph{Physical Review E} \textbf{81} (2010) 036701.

\bibitem{anteneodo2017importance}
C.~Anteneodo, S.~Camargo and R.~O. Vallejos, \enquote{Importance sampling with
  imperfect cloning for the computation of generalized {Lyapunov} exponents},
  \emph{Physical Review E} \textbf{96} (2017) 062209.

\end{thebibliography}

\end{document}